\documentclass[twocolumn,showpacs,preprintnumbers,amsmath,amssymb]{revtex4}

\usepackage{multirow}%
\usepackage{graphicx}
\usepackage{dcolumn}
\usepackage{bm}

\newcommand{\fete}{Fe$_{1.087}$Te}

\begin{document}

\title{Evidence for core-hole-mediated inelastic x-ray scattering from metallic Fe$_{1.087}$Te}

\author{J. N. Hancock$^{1}$}
\author{R. Viennois$^{1}$}
\author{D. van der Marel$^{1}$}
\author{H. M. R\o nnow$^{2}$}
\author{M. Guarise$^{2}$}
\author{P.-H. Lin$^{2}$}
\author{M. Grioni$^{2}$}
\author{M. Moretti Sala$^{3}$}
\author{G. Ghiringhelli$^{3}$}
\author{V. N. Strocov$^{4}$}
\author{J. Schlappa$^{4}$}
\author{T. Schmitt$^{4}$}

\affiliation{$^{1}$ 
D\'epartement de Physique de la Mati\`ere Condens\'ee, Universit\'e de Gen\`eve, quai Ernest-Ansermet 24, CH 1211 Gen\`eve 4, Switzerland}
\affiliation{$^{2}$\'Ecole Polytechnique F\'ed\'ederale de Lausanne, Institut de Physique des Nanostructures, CH 1015 Lausanne, Switzerland}
\affiliation{$^{3}$INFM/CNR Coherentia and Soft Ð Dipartimento di Fisica, 
Politecnico di Milano, Piazza Leonardo da Vinci 32, 20133 Milano, Italy}
\affiliation{$^{4}$Paul Scherrer Institut, CH-5232, Villigen, PSI, Switzerland}

\date{\today}

\begin{abstract}

We present a detailed analysis of resonant inelastic scattering (RIXS) from \fete\ with unprecedented energy resolution. In contrast to the sharp peaks typically seen in insulating systems at the transition metal $L_3$ edge, we observe spectra which show different characteristic features. For low energy transfer, we experimentally observe theoretically predicted many-body effects of resonant Raman scattering from a non-interacting gas of fermions. Furthermore, we find that limitations to this many-body electron-only theory are realized at high Raman shift, where an exponential lineshape reveals an energy scale not present in these considerations. This regime, identified as emission, requires considerations of lattice degrees of freedom to understand the lineshape. We argue that both observations are intrinsic general features of many-body physics of metals.
\end{abstract}

\pacs{PACS}

\maketitle
In condensed matter physics, few complete models exist to build our understanding of complex materials. Perhaps the best known paradigm in this regard is the Fermi liquid, which captures physics of diverse phenomena, from simple metals to neutron stars. A large body of theoretical work investigating the response of a Fermi liquid at an X-ray edge has uncovered diverse interesting physics of itinerant states collectively responding to a spontaneously generated core-hole potential \cite{anderson67,mahan67,nozde69,doniachsunjic69}. While the theoretical problem for the one-photon process is considered solved \cite{citrin79,doniachsunjic69}, very little is known about the two-photon response function \cite{nozab74,privalov01}. In this work, we present and interpret the resonant inelastic X-ray scattering (RIXS) cross section of FeTe, which shows both agreement and disagreement with theoretical predictions of the resonant Raman response of a noninteracting electron gas excited at an X-ray edge \cite{nozab74,mahan77,almbadh77}.

FeTe is currently intensely studied as the simplest member of a new class of high-$T_c$ superconductors \cite{hsu08,xia09,subedi08}. While recording unprecedented high quality RIXS spectra on \fete, we have discovered more general properties of the RIXS cross section which are likely to hold for essentially all metallic systems. RIXS spectra were taken at the ADRESS beamline and the SAXES spectrometer of the Swiss Light Source. The energy $\omega$ of the incident beam was tuned near the Fe $L_3$ edge (2$p$($J$=$\frac{3}{2}$)$\rightarrow$3$d$ transitions) of nonsuperconducting Fe$_{1.087}$Te. The incident energy resolution was set to $\sim$50 meV, and the total resolution of the scattered beam (energy $\omega'$) was 73 meV. Samples were grown using the Bridgeman method, all data were collected with a sample temperature near $\sim$20 K, and momentum transfer $Q$$\sim$0.51$\AA^{-1}$ directed 20$^\circ$ from the $c$ axis.

\begin{figure}
\begin{center}
\includegraphics[width=3.2in]{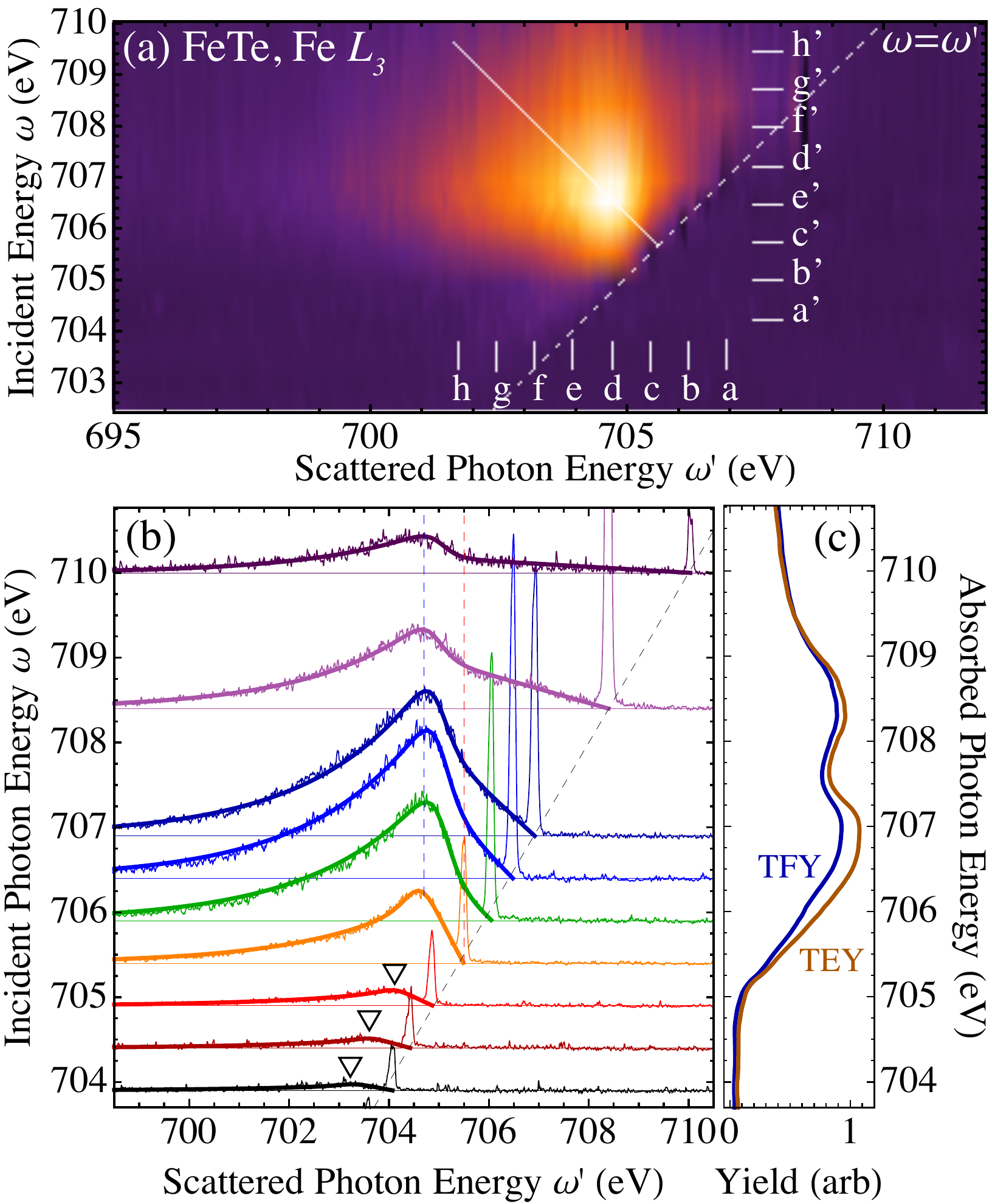}
\caption{(a) False color plot of the RIXS data, plotted versus incident and scattered photon energies, with elastic peak subtracted and linear interpolation applied between the data points. (b) RIXS spectra in detail. The solid lines show the fit to Eqn. 1, the two vertical dashed lines indicate the scatteredphoton energies $\omega'_{pk}$ (left) and $\omega'_k$ (right). Triangles point to the 0.9 eV feature in the Raman-dispersing, sub-threshold regime. (c) Total fluorescence yield (TFY) and total electron yield (TEY) spectra, showing the two-peak $L_3$ structure. }
\label{ }
\end{center}
\end{figure}

Figure 1a displays high-resolution RIXS data in the form of a contour plot with linear interpolation between the data points. The intensity is strongest when 705 eV$<$$\omega$, 
and the overall intensity appears to have even parity about the diagonal solid line, which we return to later. This resonant inelastic X-ray scattering cross section at the $L_3$ edge of Fe represents the response of a many-particle, itinerant electron system to the sudden generation of a localized potential, with the concomitant addition of 1 electron to the Fe 3$d$ subsystem. The spectral profile of these dynamics is the thesis of this paper.

Line scans are plotted versus scattered photon energy in Figure 1b. The sharpest peaks appear in each scan at the elastic line ($\omega$=$\omega'$), and demonstrate the superlative resolution of the SAXES spectrometer and ADRESS beamline. At low incident energy, $\omega$$<$705 eV, an inelastic peak (triangles) tracks the incident energy at fixed 0.9 eV loss. This Raman-Stokes-dispersing feature involves the Fe orbitals and likely results from Fe-Te hybrid excitation. As the incident energy is raised across the threshold, a peak appears at a fixed scattered photon energy $\omega'_{pk}$$\sim$704.7 eV. For the scans in this region, a kink is also clearly observed at a slightly higher energy $\omega'_k$$\sim$705.5 eV. Comparison with Figure 1c reveals that these characteristic energy scales reside near the absorption \textit{threshold}, not the maximum near 706.7 eV. 

In Figure 2a, two RIXS line scans are plotted versus energy transfer $\Delta\omega$$=$$\omega$-$\omega'$, which are typical of the above-threshold data. Each such scan shows a well-resolved peak, clearly dividing the data into two behavior types. For scans above threshold, and at low energy transfer, the intensity shows a $\Delta\omega$-linear increasing behavior, which persists to the kink. Nearly immediately above the peak, however, a fast $\Delta\omega$-falloff is observed for scans both below and above threshold. The clear change in analytic behavior of the scattering cross section when the scattered photon crosses the absorption threshold highlights the importance of  Fermi threshold phenomena in determining the RIXS signal in this metallic system. Based on this observation and the other properties of the two-dimensional spectrum, we shall see that an interpretation of these results in theory addressing edge singularities \cite{nozab74,privalov01} is appropriate.

Simply scaling of each above-threshold curve, shown in Figure 2b, can produce reasonable collapse to an exponential form, with minor inflections about this average behavior. Furthermore, a common exponential decay factor $a$ determines the lineshape of all spectra, suggesting it is a material parameter and an energy scale $a^{-1}$=2.12 eV arising from the system under study. This behavior is persistent for even high incident energy, suggesting that this portion of the RIXS response is normal X-ray emission (XES). Figure 2b inset shows that similar approximate trends are evident in the XES lineshape of varied systems such as insulating Dy(NO$_3$)$_3$ \cite{hama91} ($a^{-1}$$\sim$ 5.56eV) and Si \cite{wiech93} ($\sim$ 3.2eV), and metallic Fe \cite{yang09} ($\sim$ 1.67eV) and Li \cite{tagle80} ($\sim$1.52eV). 

This dependence of fluorescence on energy is not expected based on purely electronic considerations \cite{nozab74,eisen74}, which anticipate a weak power law falloff. However, in the case of Li, theoretical calculations by Mahan \cite{mahan77} and Almbladh \cite{almbadh77} have successfully \cite{tagle80} described the lineshape of the X-ray emission by accounting for the role of phonons in describing the profile. Two important parameters of these theories are the Auger decay rate $\Gamma_A$ and the characteristic phonon oscillation frequency $\Omega$. When $\Gamma_A>\Omega$, a nonadiabatic limit is realized, and Franck-Condon physics describe the lineshape \cite{mahan77}. With a long-lived core hole, however, the lattice can undergo several oscillations before recombination occurs and two peaks in the emission lineshape arise from deexcitation near the classical turning points of the excited state potential energy surface. In the case of the Fe $L_3$ edge of FeTe, calculations put the entire phonon spectrum below 45 meV while the core hole broadening is of order 200 meV, so we expect a dominant single peak emission spectrum. Indeed, we observe only very weak inflections about the exponential form.

\begin{figure}
\begin{center}
\includegraphics[width=3.0in]{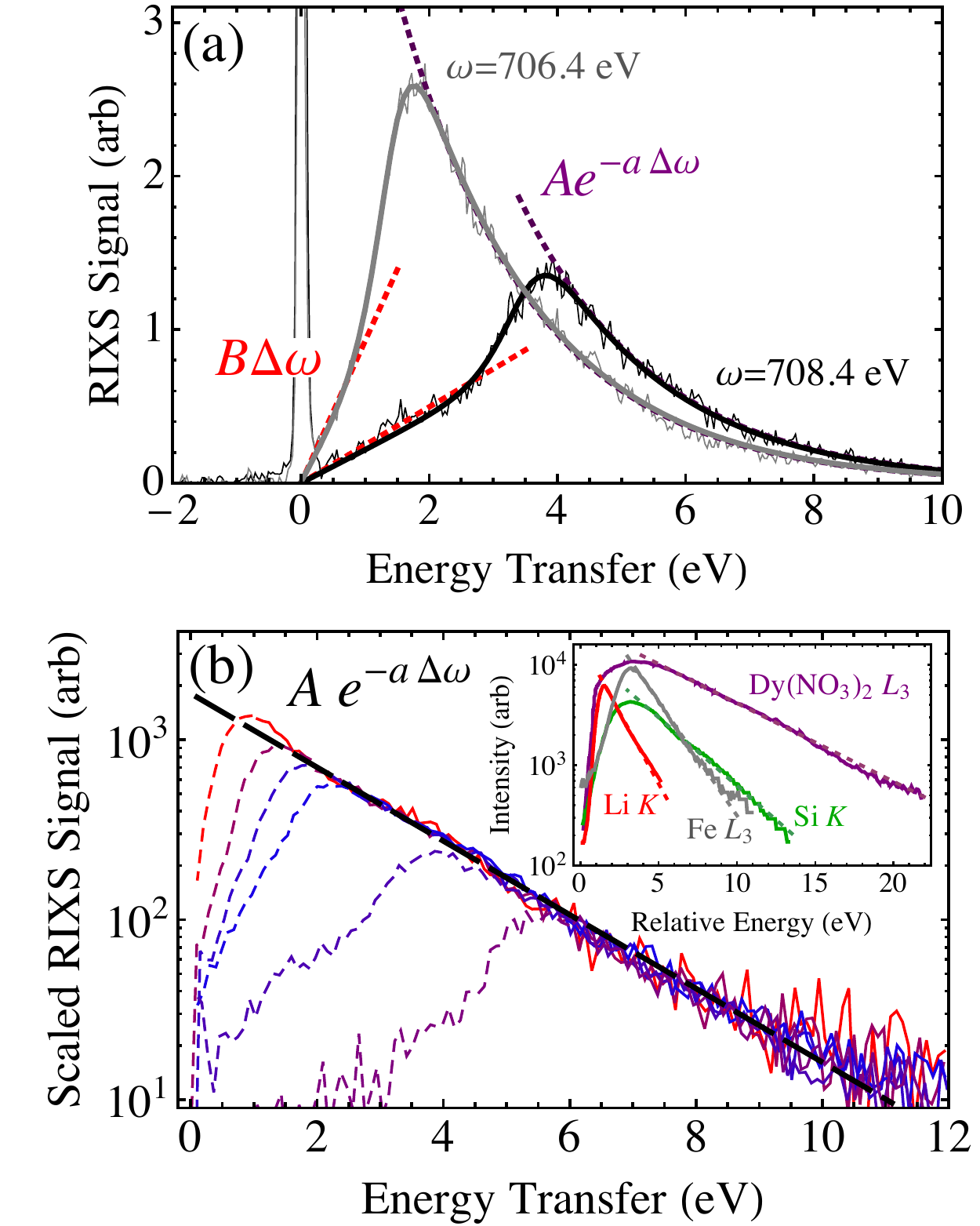}
\caption{(a) RIXS line scans $\omega$=706.4 and 708.4 eV, showing the linear and exponential behavior regimes. (b) Scaled spectra, showing the collapse at high frequencies for a single value of the decay constant $a$=0.472 eV$^{-1}$. Inset: X-ray emission spectra from the Li $K$ \cite{tagle80}, Si $K$ \cite{wiech93}, Fe $L_3$ \cite{yang09}, and Dy $L_3$ edge \cite{hama91} of Dy(NO$_3$)$_3$, measure in energy relative to the edge, showing approximate exponential behavior over a 5-15 eV range.}
\label{ }
\end{center}
\end{figure}

To quantify the incident energy-dependent trends in our data, we use a fit function that interpolates between the low-$\Delta\omega$ linear region and the high-$\Delta\omega$ exponential region, while allowing for a fast and smooth crossover. We accomplish this using the following form to represent the energy loss spectra:
\begin{equation}
\label{eq:Ifit}
I_{fit}= I_0\big[\alpha \Delta\omega (1-g_\Gamma)+ e^{-a \Delta\omega}g_\Gamma\big]
\end{equation}
where $\alpha$ represents the linear slope, $e^{-a\Delta\omega}$ the exponential tail and the function $g_\Gamma$=$(e^{-\frac{\Delta\omega-\Delta\omega^*}{\Gamma}}+1)^{-1}$ provides a smooth crossover of width $\Gamma$ at an energy $\Delta\omega^*$.

The results of fitting to this form are displayed with the data in Figure 1b and Figure 3 with $a$=0.472 eV$^{-1}$ held constant across the data set. The overall intensity parameter $I_0$ has a step-like increase above threshold because it reflects the total of all $3d$$\rightarrow$$2p$ core-hole decay channels. 
The peak energy $\Delta\omega^*$ increases linearly for above threshold scans, extrapolating to 704.9 eV, but locks to a constant value $\sim$0.9 eV for scans with $\omega$$<$705.5 eV. This behavior is typical of Raman-to-fluorescence crossover, and was discussed quantitatively in \cite{eisen74}. 
The crossover region is characterized by an energy scale $\Gamma$ (Fig 3c). The overall magnitude of the broadening is of order the core hole relaxation rate \cite{citrin79}, suggesting that the nascent nature of the core hole is an important influence in limiting the sharpness of features in the inelastic spectrum. This is contrary to what has been said of true Raman processes \cite{kotani01}, suggesting that the intensity at the peak region already feels the crossover from Raman to XES-like behavior in the exponential region. $\Gamma$ has a marked ($\sim$34\%) incident energy dependence, showing a clear sharpening of the peak near threshold, a phenomenon discussed in \cite{eisen74}.
The linear slope $\alpha$ is new, and appears to grow large as the threshold is approached from above, as is clearly visible in the raw data (Fig. 1b).

\begin{figure}
\begin{center}
\includegraphics[width=3.2in]{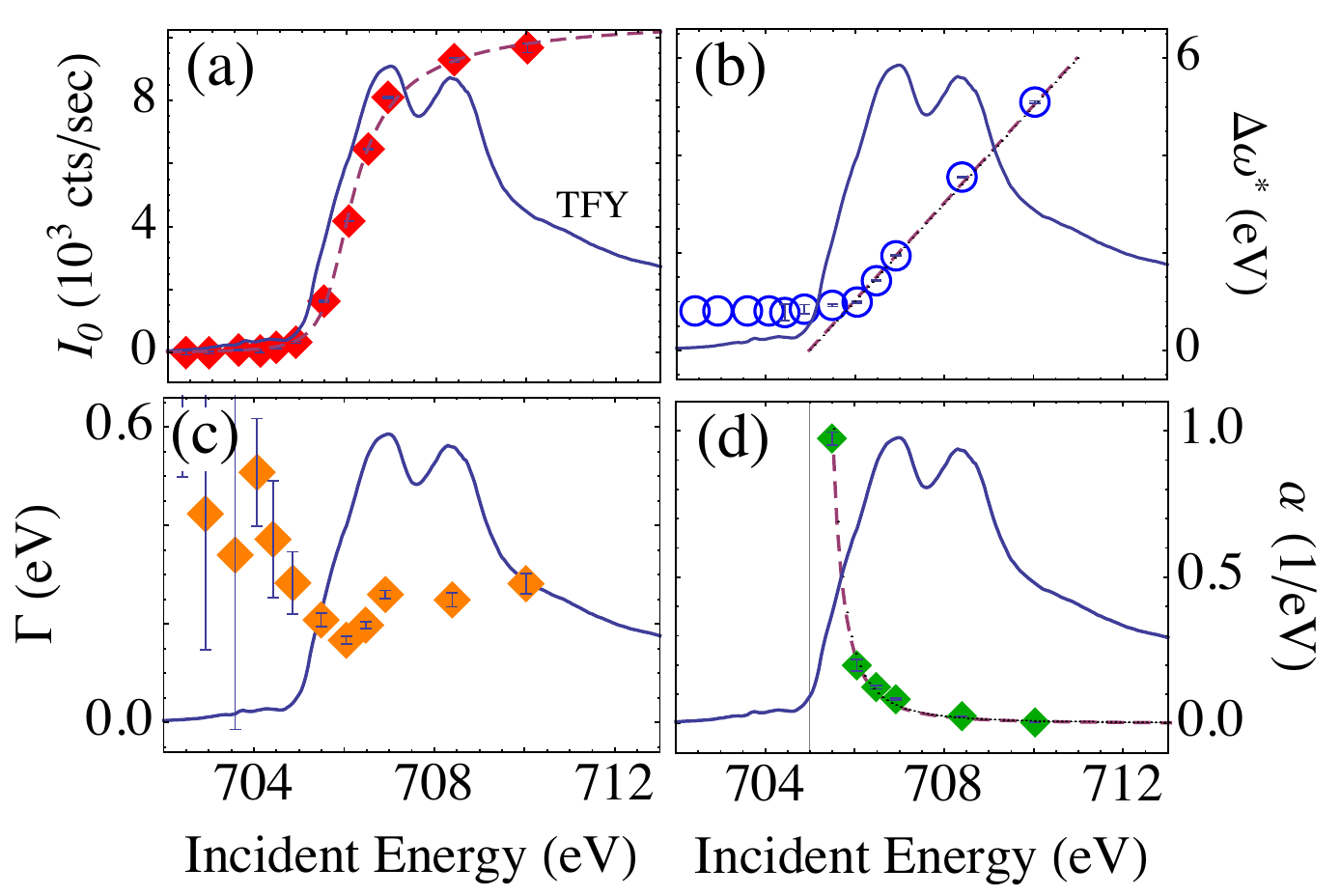}
\caption{Parameters deduced from fitting to Eqn (1), along with the TFY spectrum. In (a), the dashed line corresponds to the expression $A$(arctan$((\omega-\omega_0)/\gamma)+\pi/2)^2$. 
In (d), the dashed line corresponds to a fit to $\Delta\omega/(\omega-\omega_{th})^{2+2\eta}$ with the threshold determined from the extrapolation of the dashed line in (b). 
}
\label{ }
\end{center}
\end{figure}

To further interpret the linear slope $\alpha$, which to our knowledge has not been observed before, we evoke the prescient work of Nozi\`eres and Abrahams (N-A) which addressed the then-future possibility of realizing what we now call RIXS \cite{nozab74}.
The N-A theory explored partial solutions to the problem of a free electron gas responding to a strong local disturbance, sometimes called `final-state interactions'. These effects occur in the intermediate state of RIXS (final of absorption or photoemission), and are quite strong. In an extension of the formalism which handled the X-ray absorption problem \cite{anderson67,mahan67, nozde69,doniachsunjic69}, N-A were able to predict certain limiting behavior of the Raman scattering cross section $W_R(\omega,\omega')$. With some caveats, they predicted that for low transfer and high incident energy, $W_R(\omega,\omega')$$\sim$$\Delta\omega/(\omega-\omega_{th})^{2+2\eta}$, where $\eta$ 
is related to the radial phase shifts of the electronic states about the core-hole. By inspection of our data (Fig 1b), the linear slope increases monotonically as the energy is lowered to $\omega_{th}$$\sim$705 eV, and our fit parameter $\alpha$ quantifies this trend. Figure 3d shows a fit of the above-threshold $\alpha$ to $1/(\omega-\omega_{th})^{2+2\eta}$, keeping $\eta$=0 for simplicity, and with $\omega_{th}$ taken from the analysis in Figure 3b. We expect that band-structure effects and other non-idealized conditions will influence the applicability of this simple expression, but the qualitative behavior appears to be correctly described.



The linear dependence arises from the phase-space constraints imposed by the exclusion principle. For a wide-band metal at half filling, the density of excitations with energy $E$ and exactly $N$ electrons and $N$ holes varies as $E^{N-1}/B^N$, where $B$ is the characteristic bandwidth energy of the metal. At low energy, $E$$\ll$$B$, the $N$=1 contributions dominate and the scattering appears linear in the energy transfer. In \fete, the region of validity of this form extends as high as $\sim$4.5 eV within the errors of our measurement. This would imply a bandwidth of comparable size. Subedi, \textit{et al} calculate the band structure of \fete\ and find Fe 3$d$ bands 4-6 eV wide, consistent with the observed persistent linear behavior \cite{subedi08}. Guided by our high resolution data, where the linear slope is unambiguously observed, it also appears to be present in the data of Kurmaev \textit{et al} \cite{kurmaev09} and Yang \textit{et al} \cite{yang09}, who concluded that different metallic systems exhibit similar behavior. We therefore conclude that the salient features of the RIXS spectra we have uncovered are robust generic properties of metals.



\begin{figure}
\begin{center}
\includegraphics[width=3.45in]{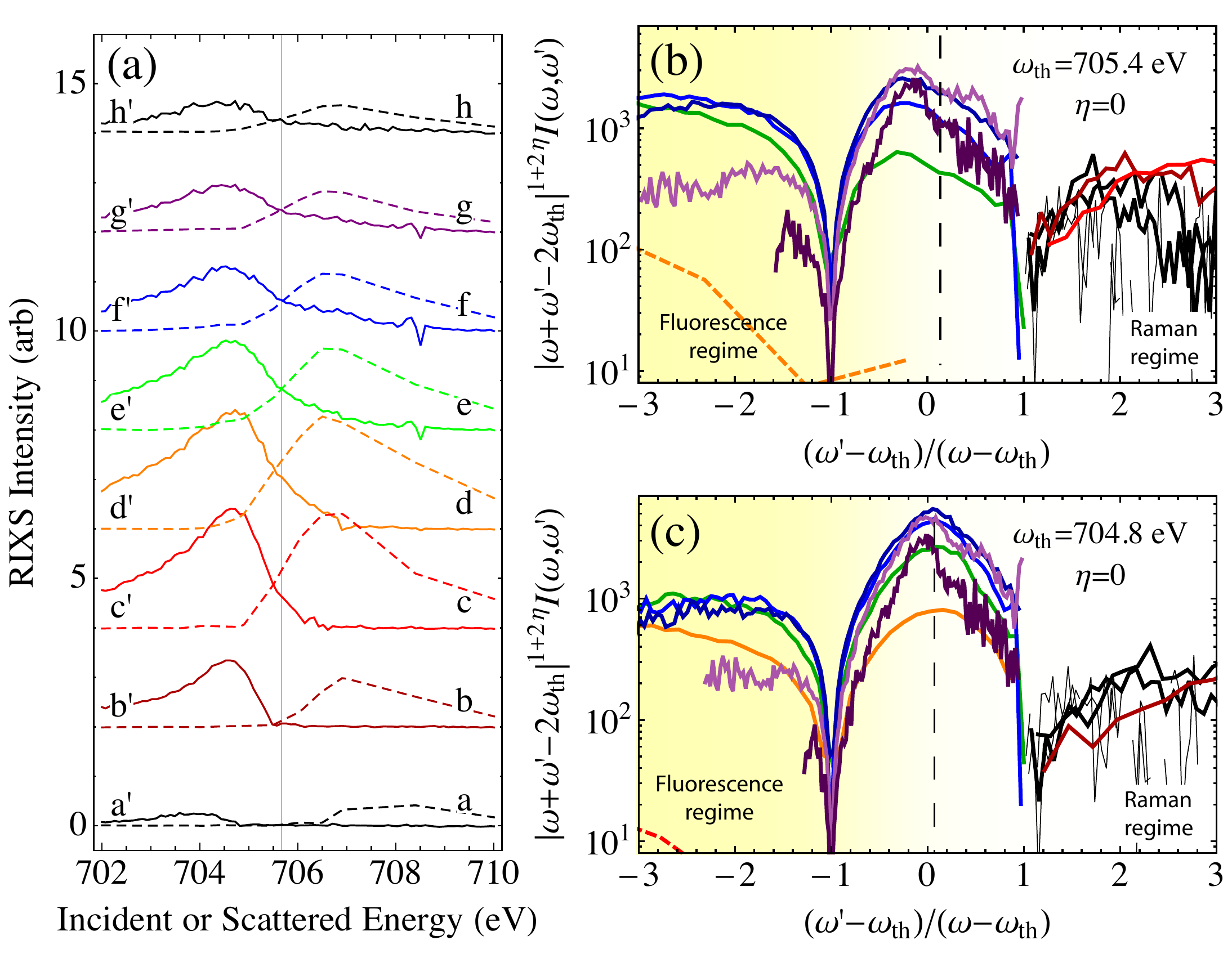}
\caption{(a) Cuts along the directions indicated in Figure 1a, versus incident (unprimed letters) or scattered (primed letters) energy. (b-c) Test of scaling relation for all measured spectra with two different threshold values. The colors are the same as in Figure 1b, with the small-signal data with $\omega<$703.9 eV in black. Dashed lines are near-threshold spectra, and not necessarily expected to obey the scaling relation.}
\label{ }
\end{center}
\end{figure}


N-A theory predicts certain symmetries to the scattering cross section, which results from the electron-hole symmetry of the single wide band metal which they chose to study. This is not strictly true in a general system, but may be approximately true for Fe 3$d$ states, which are nearly half-filled. Indeed, the electronic structure calculations of Subedi et al show a broadly distributed density of Fe 3$d$ states roughly centered on the chemical potential, motivating us to test the symmetry relation in our data. The electron-hole symmetry corresponds in RIXS to a reflective symmetry about the solid line in the false color plot of Figure 1a. Minimizing the difference $I(\omega-\omega_0,\omega'-\omega_0)$-$I(-(\omega'-\omega_0),-(\omega-\omega_0))$ on a mesh of data points determines the frequency $\omega_0$=705.6eV, within 1eV of the threshold determined by other indications in Figure 3. Figure 4a shows data along each of the cuts indicated in Figure 1a. The scans indeed look similar about this point, but we note that uncertainties in the absorption correction play a role in this comparison, limiting the accuracy to which we can test this relation.

Finally, N-A suggest that in the Raman regime, and away from threshold, the RIXS signal exhibits a scaling property:
\begin{equation}
\label{ }
W_R(\omega,\omega')|\omega+\omega'-2\omega_{th}|^{1+2\eta}=f\Big(\frac{\omega'-\omega_{th}}{\omega-\omega_{th}}\Big)
\end{equation}
Figure 4b and 4c show $f(x)$ determined for two values of the threshold energy. The Raman regime has $\omega-\omega_{th}$ and $\omega'-\omega_{th}$ of the same sign, and so corresponds to $x>0$, where reasonable collapse is observed. This collapse appears to persist partially into fluorescence regime, beyond which the scaling becomes markedly worse for $x<-1$, where the exponential tails are the most prominent. Our results thus give the first experimental insight into the form of the function $f(x)$, which appears peaked for $x$$\sim$0 ($\omega'$$\sim$$\omega_{th}$), with broad tails for large $|x|$. 
The fidelity of the scaling relation should motivate further theory to assess whether vital material-specific information can be deduced from similar analysis.

We have found in the analysis of Figures 3d and 4b,c that our data are consistent with a singularity index $\eta$ has a small value $\eta\sim0.02\pm0.09$. In the simplest considerations, $\eta=2\delta_l/\pi-\sum(2l'+1)(\delta_l/\pi)^2$ is related to the phase shifts $\delta_l$ of conduction electron spherical waves upon introduction of the core hole potential, which are constrained by the Friedel sum rule \cite{nozde69}. The sign of $\eta$ determines whether many-body effects enhance or suppress the threshold absorption \cite{doniachsond}, and depends strongly on the specific scattering process \cite{nozde69,doniachsond}. RIXS data on Li \cite{krisch97} suggest the additional importance of exchange effects \cite{girvhop76} in even simple metallic systems. For the 3$d$ transition metal $L$ edges, correlation and magnetic interactions can be very large and drive intriguing physical behavior, the importance of which to X-ray edge physics and RIXS require further work to understand. 

In summary, our results demonstrate that RIXS scattering from metallic \fete\ exhibits theoretically predicted properties which can be understood in terms of the same Fermi threshold physics which influences absorption and photoemission lineshapes. In the fluorescence regime, we have identified an energy scale which appears to reflect material properties related to lattice vibrations. Our observations are significant because they constitute the first quantitative analysis of high-resolution RIXS in a metallic system. Future theoretical and experimental work using high resolution RIXS will help clarify the relative roles of Fermi liquid parameters including electron-hole pair lifetimes, momentum transfer, and exchange phenomena in determining how a photon scatters from a metal near an X-ray absorption edge.

We would like to acknowledge valuable conversations with E. Abrahams, T. Giamarchi, G.-H. Gweon, P. M. Platzman, and B. S. Shastry and support by Swiss National Science Foundation, Materials with Novel Electronic Properties (MaNEP). ADRESS/SAXES was developed jointly by Politecnico di Milano, SLS, and EPFL.

\bibliography{rixs}
\end{document}